\documentstyle[12pt]{article}
\def\beq{\begin{equation}}
\def\eeq{\end{equation}}

\begin{titlepage}
\begin{center}
{\Large \bf Theoretical Physics Institute \\
University of Minnesota \\}  \end{center}
\vspace{0.3in}
\begin{flushright}
TPI-MINN-92/50-T \\
September 1992
\end{flushright}
\vspace{0.4in}
\begin{center}
{\Large \bf Summing one-loop graphs in a theory with broken symmetry\\}
\vspace{0.2in}
{\bf B.H. Smith  \\ }
School of Physics and Astronomy \\
University of Minnesota \\
Minneapoilis, MN 55455 \\

\vspace{0.2in}
{\bf   Abstract  \\ }
\end{center}

The method suggested by Lowell Brown for calculating multi-particle
threshold amplitudes is extended to the one-loop level in scalar theories
with broken reflection symmetry.  A result for the threshold amplitude for
multiparticle production is derived.  It is also shown that the tree-level
amplitude for 2 on-mass-shell particles producing $n$ particles vanishes at
the kinematical threshold for $n>2$.  
\end{titlepage}

\begin{document}

There has been a great deal of interest in processes involving the
production of many particles in weakly interacting theories. In the past
year, the tree-level contribution for $1 \to n$ processes at the kinematical
threshold has been solved exactly for $\lambda\phi^4$ scalar theory using
recursion relations$^{\cite{voloshin1,akp1}}$.  This threshold computation
serves as an upper limit for estimates of the nonthreshold amplitude and can
be used to calculate a lower limit$^{\cite{voloshin2,akp2}}$.

Recently, Brown$^{\cite{brown}}$ has demonstrated an equivalent method for
summing all tree-level Feynman diagrams by solving the classical equations
of motion.  This method elegantly yields the same results as the previous
papers using recursion relations$^{\cite{voloshin1,akp1}}$. Brown has solved
for the amplitude in $\lambda\phi^4$ theories with both broken and unbroken
reflection symmetry ($\phi \to -\phi~$).

It has been shown by Voloshin$^{\cite{voloshin3}}$ that this method can be
extended to find the threshold contribution of one-loop processes. Voloshin
has calculated the threshold amplitude for the case of unbroken symmetry.
The threshold amplitude for one off shell scalar particle going into $2k+1$
scalar particles was calculated to be

\beq
\langle 2k+1|\phi(0)|0\rangle = (2k+1)! \left ( {\bar\lambda\over{8
\bar{m}^2}}\right )^k \left [1-k(k-1){3\sqrt{3}\lambda\over{16\pi^2}} \left
(\ln{{2+\sqrt{3}}\over{2-\sqrt{3}}}-
i \pi \right ) \right ],
\label{unbrokengreen}
\eeq

where $\bar{\lambda}$ is the renormalized coupling constant and $\bar{m}$ is
the renormalized mass. The number of produced particles, $2k+1$, is
neccesarily odd to preserve the reflection symmetry.  In doing this
calculation, an interesting phenomena was discovered. Voloshin proved that
the threshold amplitude for 2 on-shell particles going into $n$ particles
vanishes in $\lambda\phi^4$ theory with unbroken reflection symmetry.

In this paper, we show that this method can be extended to theories with
broken symmetry.  In such theories, we calculate the first loop correction
to the amplitude. The result is

\beq
\langle n|\phi(0)|0\rangle = n! \left ( {\bar\lambda\over{2
\bar{M}^2}}\right )^{(n-1)/{2}} \left (1+n(n-1){\sqrt{3}\lambda\over{8\pi}}
\right
),
 \label{brokengreen}
\eeq
where $\bar{\lambda}$ is the renormalized coupling constant and $\bar{M}$ is
the renormalized mass of the physical scalar boson as defined in eq.
($\ref{renormalizedValues}$).

In addition, we show that for the theory with broken symmetry, the tree
level amplitude of 2 on-shell scalar particles going into $n$ particles
vanishes at the kinematical threshold for $n>2$.  \\ \\

In this paper we are considering $\lambda\phi^4$ theory with broken symmetry
which has the Lagrangian

\beq
{\cal L} = {1\over{2}}(\partial\phi)^2 + {1\over{2}}m^2\phi^2 -
{\lambda\over{4}}\phi^4.
\label{unblangrange}
\eeq
In this theory, the field $\phi$ develops a vacuum expectation value of
$\langle\phi\rangle = \sqrt{m^{2}/\lambda}$. We use the shifted field
$\sigma \equiv \phi - \langle\phi\rangle$ with the Lagrangian

\beq
{\cal L} = {1\over{2}}(\partial\sigma)^2 - {1\over{2}}M^2\sigma^2 -
M\sqrt{\lambda\over{2}}\sigma^3-{\lambda\over{4}}\sigma^{4},
 \label{blangrange}
\eeq
where $M = \sqrt{2}m$ is the effective mass of the shifted field, $\sigma$.

Brown's approach$^{\cite{brown}}$ uses the reduction formula method to
generate the amplitude by coupling the field, $\sigma$, to a source,
$\rho(x)$, in the form of an additional term in the Lagrangian,
$\rho\sigma$.  The amplitude is then written as

\beq
\langle n|\sigma|0\rangle = \left [\prod_{a=1}^n \lim_{p_a^2 \to m^2}\int
d^4 x_a e^{i p_a x_a} \left (M^2-p_a^2\right )
{\delta\over{\delta\rho(x_a)}}\right ] \langle
0+|\sigma(x)|0-\rangle^{\rho}|_{\rho=0}.
\label{reducform}
\eeq
The response of the field, $\langle 0+|\sigma|0-\rangle$, can be
approximated at the one loop level by the mean value of the field,
$\sigma(x)=\sigma_{cl}(x)+\sigma_{1}(x)$, where $\sigma_{cl}(x)$ is the
solution to the classical field equations, and $\sigma_{1}(x)$ is the
average value of the quantum part of the field.

Since all the spatial momenta of the final particles vanish, the threshold
source and response functions can be considered as spatially independent
functions, $\rho(t)$ and $\sigma(t)$.  The response of the free field in the
presence of the source, $z(t)$, can easily be found by solving the equations
of motion in the limit of vanishing $\lambda$. The resulting solution is

\beq
z(t) = {\rho(t)\over{M^2-\omega^2-i\epsilon}} = {\rho_0 e^{i\omega
t}\over{M^2-\omega^2-i\epsilon}.
\label{zdef}}
\eeq
Eq. ($\ref{zdef}$) can be used the rewrite eq. ($\ref{reducform}$) in a
simple form that obviously remains finite when the limits as $\omega \to M$
and $\rho_0 \to 0$ are taken simultaneously. The expression for the
amplitude will then become

\beq
\langle n|\sigma|0\rangle = \left ( {\partial\over{\partial z}}\right )^n
(\sigma_{cl}+\sigma_{1})|_{z=0}.
 \label{genfun}
\eeq

As $\rho_0 \to 0$ the classical field, $\sigma_{cl}(t)$, satisfies the
equation

\beq
{\partial^2 \sigma_{cl}\over{\partial t}^2} + M^2 \sigma_{cl} +3
M \sqrt{\lambda\over{2}}\sigma_{cl}^2 + \lambda\sigma_{cl}^3 = 0.
 \label{classfield}
\eeq
The classical response should reduce to the free field response, z(t), in
the limit as $\lambda \to 0$. The solution that satisfies both of these
conditions is$^{\cite{brown}}$:

\beq
\sigma_{cl}(t) ={ {z(t)}\over{1-z(t)\sqrt{\lambda/{2M^2}}}}
\label{classicalresponse}
\eeq

We now turn to calculate the quantum correction. We can expand the field
about the classical solution and use eq. ($\ref{classfield}$) to find that
$\sigma_{1}$ must satisfy

\begin{eqnarray}
{\partial^2 \sigma_{1}(t)\over{\partial t}^2} + M^2 \sigma_{1}(t) +6
M\sqrt{\lambda\over{2}}\sigma_{cl}\sigma_{1}(t) +
3\lambda\sigma_{cl}^2\sigma_{1}(t)+  \nonumber \\
3M\sqrt{\lambda\over{2}}\langle\sigma_q(x)\sigma_q(x)\rangle  +
3\lambda\sigma_{cl}\langle\sigma_q(x)\sigma_q(x)\rangle = 0.
 \label{quantfield}
\end{eqnarray}
where $ \langle \sigma_q(x)\sigma_q(x)\rangle $ is the $x \to x'$ limit of
the Green function $G(x,x')  \equiv \langle
T\sigma_{1}(x)\sigma_{1}(x')\rangle$. This Green function is the inverse of
the second variation of the Lagrangian,

\beq
\left [\partial^2 + M^2 + 6
M \sqrt{\lambda\over{2}}\sigma_{cl}(t)+3\lambda\sigma_{cl}(t)^2 \right ]
G(x,x') = -i\delta(t-t')\delta^3({\bf x-x'})
\label{greenform}
\eeq
To solve eq. ($\ref{greenform}$) for this Green function, we will first solve
for the solutions of the homogeneous equation with the operator on the left
hand side of eq.
($\ref{greenform}$). The two solutions, one regular as $t \to +\infty$ and
the other as $t \to
-\infty$, can be combined to form the Green function.

This problem can be rendered more tractable by analytic continuation into
Euclidean space. We
can keep the poles off the real $\tau$ axis by relacing $z(t)$ with
$u(\tau)\equiv e^{M\tau}$
defined so that

\beq
u(\tau) = -\sqrt{\lambda\over{2M^2}}z(t).
\label{udef}
\eeq
This fixes the variable $\tau$ to be

\beq
\tau = it + {1\over{2M}}\ln{{\lambda z(0)^2}\over{2M^2}}
\label{taudef}
\eeq

In terms of $u(\tau)$, $\sigma_{cl}$ (eq. ($\ref{classicalresponse}$))
can be rewritten as

 \beq
\sigma_{cl}(u(\tau)) = {-\sqrt{2M^2/{\lambda}}u(\tau)\over{1+u(\tau)}}.
\label{simgaOfTau}
\eeq
This allows us to recast the operator in eq. ($\ref{greenform}$) in terms of
hyperbolic trigonometric functions. In this way the Green function can be
found by solving

\beq
\left [ {\partial^2\over{\partial\tau^2}}-\omega^2+{3 M^2\over{2
\cosh^{2}{M\tau/2}}} \right ] f(\tau) = 0
 \label{halfgreen}
\eeq
where $\omega^2={\bf k}^2+M^2$, and ${\bf k}$ is the conserved spatial
momentum of a given
mode. This is equivalent to the well known problem worked out in Quantum
Mechanics$^{\cite{landau}}$. The solutions of eq. ($\ref{halfgreen}$)
turns out to be
\beq
f_{-}(u(\tau))= {2 u^{\omega/M} (
M^2-4M^2 u+M^2
u^2+2M\omega-3M^2u^2\omega+2\omega^2+4u\omega^2+
2u^2\omega^2)\over{M^2(1+u)^2}}
\label{fsolminus}
\eeq
regular as $\tau \to -\infty$ and

\beq
f_{+}(u(\tau))= {2 (
M^2-4M^2 u+M^2
u^2-2M\omega+3M^2u^2\omega+2\omega^2+4u\omega^2+2u^2\omega^2)
\over{u^{\omega/M}M^2(1+u)^2}}
\label{fsolplus}
\eeq
regular as $\tau \to +\infty$. The Wronskian of these solutions is
\beq
W = f_{+}(\tau)f_{-}^{'}(\tau)- f_{+}^{'}(\tau)f_{-}(\tau) =
{8\omega(\omega^2-M^2) (4\omega^2-M^2)\over{M^4}}.
\label{wronskian}
\eeq

We are now in a position to solve for the frequency components of the
two-point Green function. By looking at
$G_{\omega}(\tau_{1},\tau_{2})=f_{+}(\tau_{1})f_{-}(\tau_{2})/W$ for
$\tau_{1}>\tau_{2}$ and
$G_{\omega}(\tau_{1},\tau_{2})=f_{+}(\tau_{2})f_{-}(\tau_{1})/W$ for
$\tau_{2}>\tau_{1}$, we can extract some information about $2 \to n$
processes. Looking at the reduction formula, we have poles when $\omega=M$
or $\omega={M/{4}}$. This tells us that threshold amplitudes vanish for $2
\to n$ processes except for when the incoming particles have an energy of
M/2 or M. Since neither $2 \to 1$ or $2 \to 2$ processes have any phase
space, the rate of threshold multiparticle production with two incoming on
mass shell particles will always vanish.

In equation ($\ref{quantfield}$) one needs to look at the equal time Green
function,

\beq
G_{\omega}(\tau,\tau) = {1\over{2\omega}} + {3 M^2 u\over{2 (1+u)^{2}
\omega^3}} +{3
M^4 u[-(u+1)^2 M^2+ (1+14u+u^2)\omega^2]\over{2(1+u)^4\omega^3
(\omega^2-M^2) (4\omega^2-M^2)}},
\label{greenw}
\eeq
summed over the modes with all values of ${\bf k}$. This is done by
computing the integral

\beq
\langle \sigma_q(x)\sigma_q(x)\rangle  =\int
G_{\omega}(\tau,\tau) {d^{3}k\over{2\pi^3}} =
{1\over{2\pi^2}}\int_{M}^{\infty} G_{\omega}(\tau,\tau)
\omega\sqrt{\omega^2 -M^2}
d\omega.
\label{greenintegral}
\eeq
Looking at eq. ($\ref{greenw}$), we see that the first two terms in the
integral will be ultravioletly divergent. Other than this divergence, the
integrand is regular over the interval of integration. We can write the
divergent integrals in terms of

\beq
I_{n} \equiv {1\over{2 \pi^2}}\int_{M}^{\infty}
\omega^{1-n}\sqrt{\omega^2-M^2}  d\omega
{}~.
\label{idef}
\eeq

With a little persistence, the integral in eq. (\ref{greenintegral}) can be
computed to find the sought after Green function,

\beq
\langle \sigma_q(x)\sigma_q(x)\rangle = {I_1\over{2}} + {u
M^2\over{(1+u)^2}}\left [ {3 I_{3}\over{2}}
+{3\over{4\pi^2}}-{\sqrt{3}\over{8 \pi}} \right ] + {\sqrt{3} u^2 M^2
\over{2\pi (1+u)^4}}.
\label{greensoln}
\eeq

The divergences in  eq. ($\ref{greensoln}$) can be removed by appropriate
renormalization of the coupling constant, $\lambda$, and the mass. If the
divergent terms are written in terms of the unshifted field, $\phi_{cl}
\equiv \sigma_{cl} + \sqrt{M^{2}\over{2\lambda}}$, then it is easy to see
the renormalized values can be defined as

\begin{eqnarray}
\bar{M}^2 = M^2 -{3\lambda I_{1}}-{3\lambda\sqrt{2}M^{2}\over{2}} \left [ {3
I_{3}\over{2}} +{3\over{4\pi^2}} - {\sqrt{3}\over{8\pi}} \right ] \nonumber
\\ \bar{\lambda} = \lambda - {3\lambda^{2}\over{8}} \left [ {3
I_{3}\over{2}} +{3\over{4\pi^2}} - {\sqrt{3}\over{8\pi}} \right ]
\label{renormalizedValues}
\end{eqnarray}

Knowing the Green function, eq. ($\ref{quantfield}$) can now be cast in
terms of a solvable second order differential equation,
\beq
u {\partial\over{\partial u}}\left ( u{\partial\sigma_{1}\over{\partial
u}}\right )-\sigma_{1}+{6u\sigma_{1}\over{1+u}}-{6u^2
\sigma_{1}\over{2(1+u)^2}} = {{3\over{2\pi}}\sqrt{3\lambda\over{2}}{M
u^2\over{(1+u)^4}}}\left ( 1 - {2u\over{1+u}} \right ).
\label{sigmaQeqn}
\eeq
One can verify that eq. ($\ref{sigmaQeqn}$) will be satisfied by the
solution
\beq \sigma_{1} = + \sqrt{3\lambda\over{2}} {M
u^2\over{2\pi(1+u)^3}}.
\eeq
We can now transform our solution back into
Minkowski space to arrive at the solution for the one-loop response
\beq
\sigma = {z(t)\over{1-z(t)\sqrt{\bar{\lambda}\over{2 M^{2}}}}} \left ( 1 -
{\lambda^{3/{2}}\sqrt{3}z(t)\over{4\sqrt{2}\pi
M(1-z(t)\sqrt{\lambda\over{2M^{2}}})^2}} \right ).
\label{sigmasoln}
\eeq
When eq.($\ref{sigmasoln}$) is expanded in powers of $z(t)$,
eq.($\ref{brokengreen}$) can be easily verified.  \\ \\

The one-loop
correction for the case of broken symmetry ($\ref{brokengreen}$) is similar
in form, but opposite in sign to the case of unbroken symmetry calculated by
Voloshin$^{\cite{voloshin3}}$ ($\ref{unbrokengreen}$).  Both correction
terms grow  like $n^2$ at large $n$. However, the correction in the case of
unbroken symmetry does not aquire an imaginary part. This can be traced to
the vanishing of the two-point Green function for all real processes.

The
vanishing of the two point green function at the kinematical threshold for
$n>2$ can easily be verified for low $n$. When one carries this out, it is
seen that the vanishing can be traced to the addition of the three-point
vertices with the appropriate coupling constant. While diagrams of one sign
will have decreasing weight due to the propagator factors, the symmetry
factors and extra diagrams conspire to cause the amplitude to disappear.
Such effects may produce other yet to be discovered phenomena.  \\ \\

I would like to thank Mikhail Voloshin for bringing this problem to my
attention and for many helpful discussions. This work was supported, in
part, by Department of Energy grant DOE-DE-AC02-83ER40105.

\end{document}